\documentclass[11pt,a4paper]{article}
\usepackage[left=15mm, top=10mm, right=15mm, bottom=20mm, nohead=true, nofoot=true]{geometry}

\usepackage[utf8]{inputenc}
\usepackage{authblk}
\usepackage{indentfirst}
\usepackage{misccorr}
\usepackage{graphicx}
\usepackage{amsmath}
\usepackage{amssymb}
\usepackage{multicol}
\usepackage{bm}
\usepackage{longtable}
\usepackage{pdflscape}
\usepackage{epstopdf}
\usepackage{multirow}
\usepackage{adjustbox}
\usepackage{amsfonts}
\usepackage{ifthen}
\usepackage{fancyhdr}
\usepackage{setspace}
\usepackage{xcolor}
\usepackage[round,authoryear,comma,sort&compress,numbers]{natbib}
\usepackage[toc,title,page]{appendix}
\usepackage[hidelinks]{hyperref}
\usepackage{url}

\hypersetup{
    colorlinks=true,
    linkcolor=green,
    citecolor=blue,
    filecolor=magenta,
    urlcolor=cyan,
    pdftitle={Sharelatex Example},
    pdfpagemode=FullScreen,
}

\fancypagestyle{plain}{\chead{\texttt{\textcolor{brown}{}}}}
\title{\textbf{IdentYS: A Python-Based Tool for Identifying Young Stars in Star-Forming Regions}}
\author[]{E. Nikoghosyan \thanks{elena@bao.sci.am, Corresponding author}}
\author[]{D. Baghdasaryan}
\author[]{D. Andreasyan}
\author[]{N. Azatyan}
\author[]{A. Samsonyan}
\author[]{A. Yeghikyan}
\affil[]{\scriptsize Byurakan Astrophysical Observatory, 0213, Aragatsotn reg., Byurakan vil., Armenia}

\begin{document}
\pagestyle{empty}
\newpage
\pagestyle{fancy}
\label{firstpage}
\date{}
\maketitle

\begin{abstract}
Research on young stellar populations is essential to understand the properties of embedded clusters and advance theories of their formation. This has driven advancements in methodologies for star detection, leading to the development of valuable databases and software. We present the scientific justification and operating principles of the \texttt{IdentYS} tool, which is designed to identify young stellar objects (YSOs) in star-forming regions. The tool facilitates the identification of young stars with infrared (IR) excess in remote and embedded star-forming regions, focusing primarily on Class I and II YSOs. For this purpose, near- and mid-IR photometric data and five colour-colour diagrams (J - H) vs. (H - K), K - [3.6] vs. [3.6] - [4.5], [3.6] - [4.5] vs. [5.8] - [8.0], [3.6] - [4.5] vs. [8.0] - [24], and [3.4] - [4.6] vs. [4.6] - [12] are used. The purity of the YSOs sample is enhanced by excluding field contamination from stellar and extragalactic objects. As a result, we compile a list of YSO candidates displaying the source designation, astrometric, and photometric parameters, as well as information on the evolutionary stage determined by the presence of IR excess, as indicated by certain diagrams. The application of this program can greatly streamline the statistical analysis of young stellar populations across diverse star-forming regions, including distant and deeply embedded ones, which typically require processing large volumes of initial data.
\end{abstract}
\emph{\textbf{Keywords:} stars: pre-main sequence - stars: formation - infrared: stars - astronomical databases: miscellaneous}

\section{Introduction}
\label{introduction}

Stellar objects undergo a complex evolutionary process, beginning with their initial collapse within a progenitor molecular cloud and continuing through to the main sequence (MS) stage. During this progression, the observational characteristics of young stellar objects (YSOs) extend across the electromagnetic spectrum, from X-rays to radio waves, shaped by the presence of circumstellar envelopes, accretion disks, stellar winds, and collimated outflows. Consequently, these properties are essential for both the identification and detailed study of YSOs. At the earliest stages, including the prestellar and protostellar phases, radiation is emitted predominantly in the submillimeter and far-infrared (FIR) ranges \citep{Andre1993, Elia2017}. As the object evolves into the Class I (star+disk+envelope) and Class II (star+disk) stages, the peak of the spectral energy distribution shifts towards shorter wavelengths in the mid- and near-infrared (MIR and NIR) ranges \citep[][and ref. therein]{Allen2007}. This shift is caused by the presence of an optically thick circumstellar disk and envelope, which absorb and re-emit radiation from the central protostar, producing a distinctive observational signature of YSOs, namely an infrared (IR) excess. Class II objects display additional signs of youth, including enhanced chromospheric activity, stellar outflows, and emission from accretion shocks, where matter from the circumstellar disk is channeled along magnetic field lines onto the stellar surface \citep[][and ref, therein]{Briceno2007}. Key indicators for identifying Class II YSOs include H$_{\alpha}$ and Ca\,II H \& K emissions, 6708\,\AA\, Li\,I absorption, ultraviolet excess, and strong X-ray emission, with L$_X/L_{bol}$ ratios ranging from 0.1\% to 0.01\% \citep{Feigelson2007}. Notably, strong X-ray emission is particularly prevalent in pre-main sequence (PMS) stars at the Class III evolutionary stage, where activity is primarily driven by enhanced solar-type magnetic activity.

The properties outlined above, often combined with proper motions, are crucial for identifying young stellar population at various evolutionary stages within star-forming regions and young stellar clusters. In turn, studying the young stellar population is essential for determining the physical properties of embedded young clusters, such as their morphology, size, and stellar density. Such investigations are fundamental in improving our understanding of cluster formation processes. Consequently, scientific methodologies for detecting stars at various evolutionary stages using observational data have advanced significantly. This ongoing effort has led to the compilation of databases and the development of specialized software packages. Below, several noteworthy findings have emerged from these technological and methodological advancements are presented.

Based on data from the five FIR bands, ranging from 70 to 500\,$\mu$m, obtained as part of the Herschel Infrared Galactic Plane Survey (Hi-GAL), a catalogue of prestellar and protostellar objects has been compiled \citep{Elia2017}. 

The successful identification of Class II YSO candidates was achieved based through H$_{\alpha}$ emission analysis as part of the INT Photometric H$_{\alpha}$ Survey \citep[IPHAS,][]{Witham2008}, and spectral data from the Large Sky Area Multi-Object Fiber Spectroscopic Telescope \citep[LAMOST;][]{Tan2024}. Furthermore, catalogues of stellar members within 20 OB-dominated young clusters, all located at distances D $\leq$ 4\,kpc, were compiled as part of the Massive Young Star Forming Complex Study in the Infrared and X-ray project \citep[MYStIX;][]{Feigelson2013}.

Various software tools that employ machine learning techniques have been developed to identify YSO candidates. For example, the Spectrum Classifier of Astronomical Objects has been used to distinguish between MS stars, galaxies, and YSOs \citep{Chiu2021}. \citet{Wilson2023} introduced a Naive Bayes Classifier to identify Class II YSOs using photometric data spanning from the optical to the MIR. Other studies have also applied machine learning techniques to YSO classification \citep[e.g.][]{Marton2019, Vioque2020}. Methodologies for classifying Class I and II YSOs using Spitzer Space Telescope data are described in \citet{Kuhn2021} and \citet{Cornu2021}. In large-scale studies of the Milky Way, such as the SPICY catalogue \citep{Kuhn2021}, Gaia data are also employed to assess whether sources are associated with specific star-forming regions. This approach, however, is limited by distance, as Gaia becomes increasingly ineffective beyond 4\,–\,5\,kpc \citep{Lindegren2021}. Additionally, the significant interstellar extinction in embedded star-forming regions often makes optical data unsuitable for detailed analysis.

In this paper, we introduce the \texttt{IdentYS}, a code developed to efficiently identify potential members of active star-forming regions \footnote{The \texttt{IdentYS} code and usage instructions are available at \url{https://github.com/DanielBaghdasaryan/identys}}. The method rests on two main assumptions: (i) the vast majority of members within active star-forming regions are YSOs; and (ii) the most YSOs observed in the direction of a given region are physically associated with it. Since IR excess is one of the primary observational indicators of YSOs, our approach targets stellar objects exhibiting IR excess within the boundaries of the selected regions. The selection procedure focuses on YSOs at the Class I and Class II evolutionary stages and is applicable to both embedded and distant star-forming environments. Designed for accessibility and speed, \texttt{IdentYS} enables rapid processing of large multi-wavelength datasets, making it a practical and user-friendly tool for broad-scale stellar population studies.

The paper is organized as follows. Section \ref{Science} outlines the scientific rationale for the dataset used in our analysis. Section \ref{code} provides a detailed description of the Python code. Section~\ref{results} presents the results of the identification of YSOs within a certain star-forming region. Finally, Section~\ref{Summary} summarizes the main findings of the study.

\begin{figure*}
	\centering 
	\includegraphics[width=0.46\textwidth, angle=0]{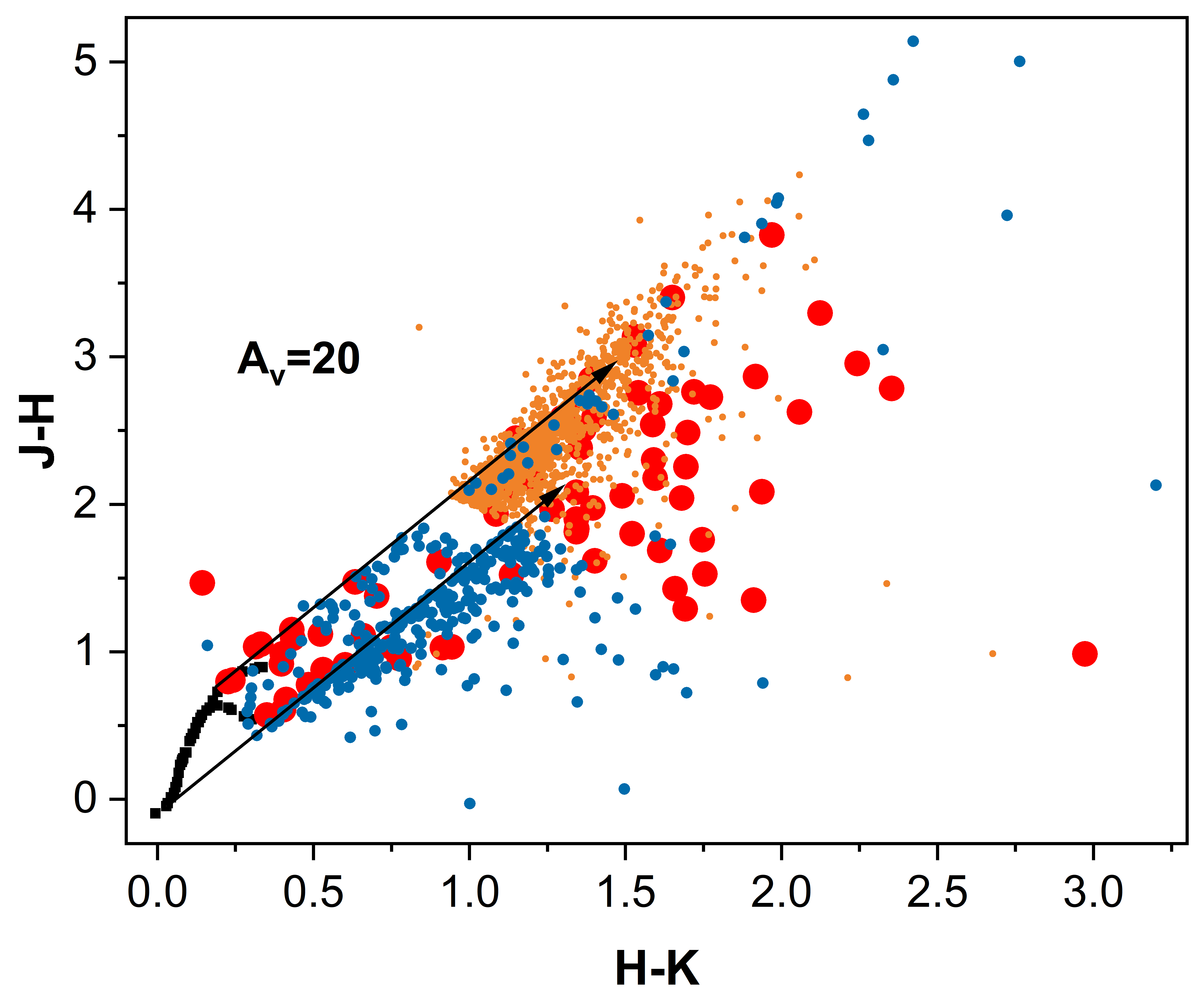}	
 \includegraphics[width=0.48\textwidth, angle=0]{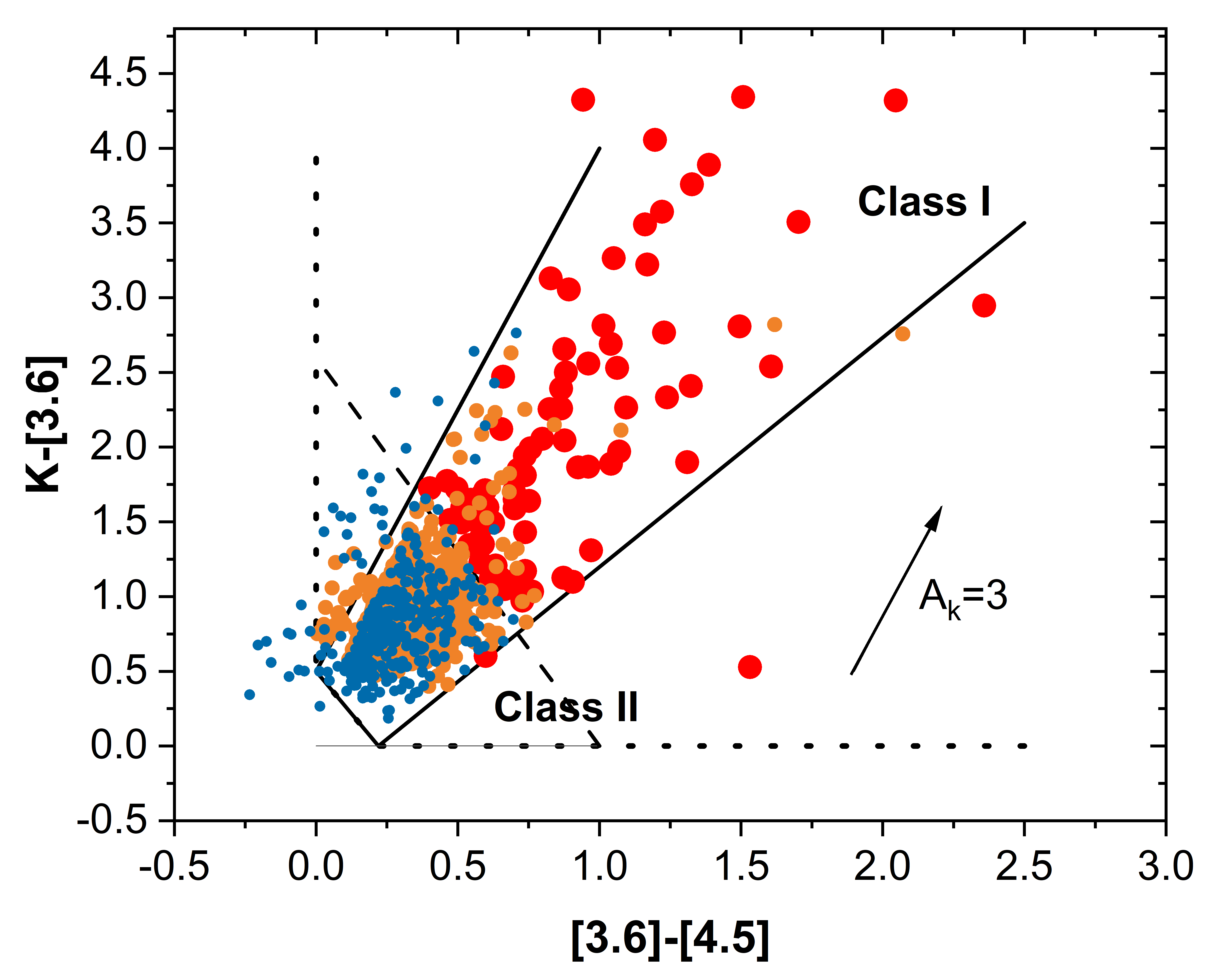}	
 \includegraphics[width=0.46\textwidth, angle=0]{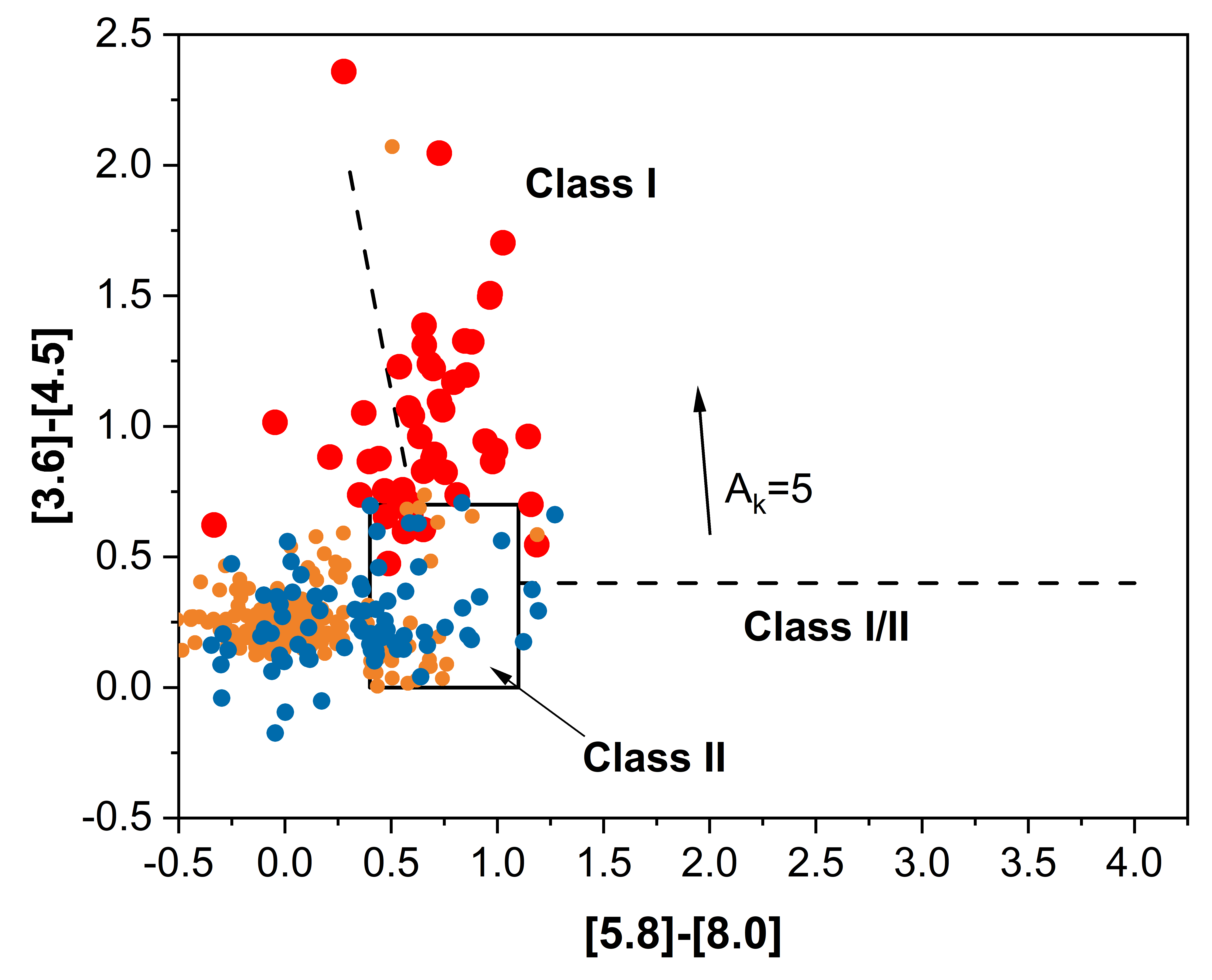}
 \includegraphics[width=0.45\textwidth, angle=0]{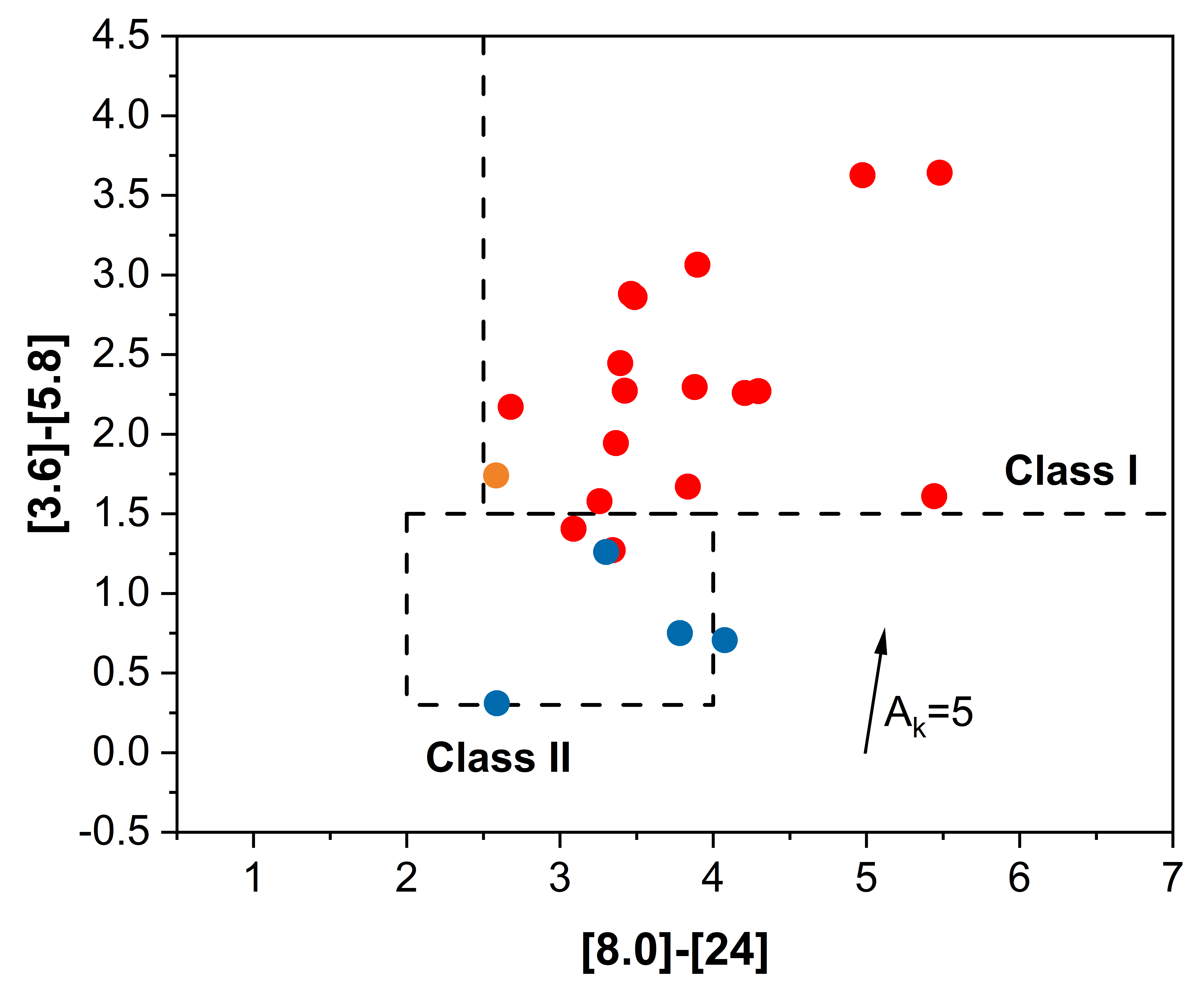}
 \includegraphics[width=0.46\textwidth, angle=0]{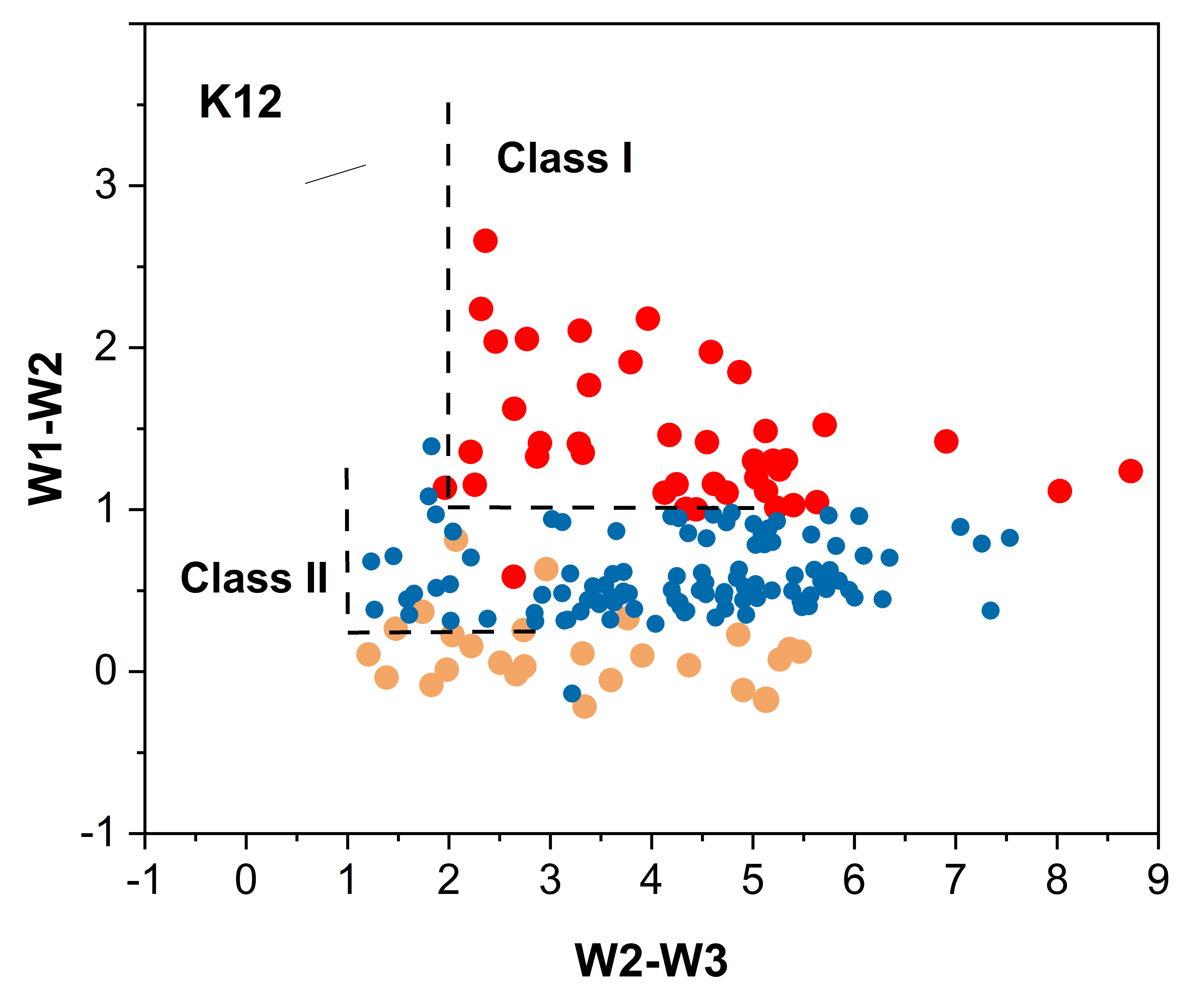}
  \includegraphics[width=0.46\textwidth, angle=0]{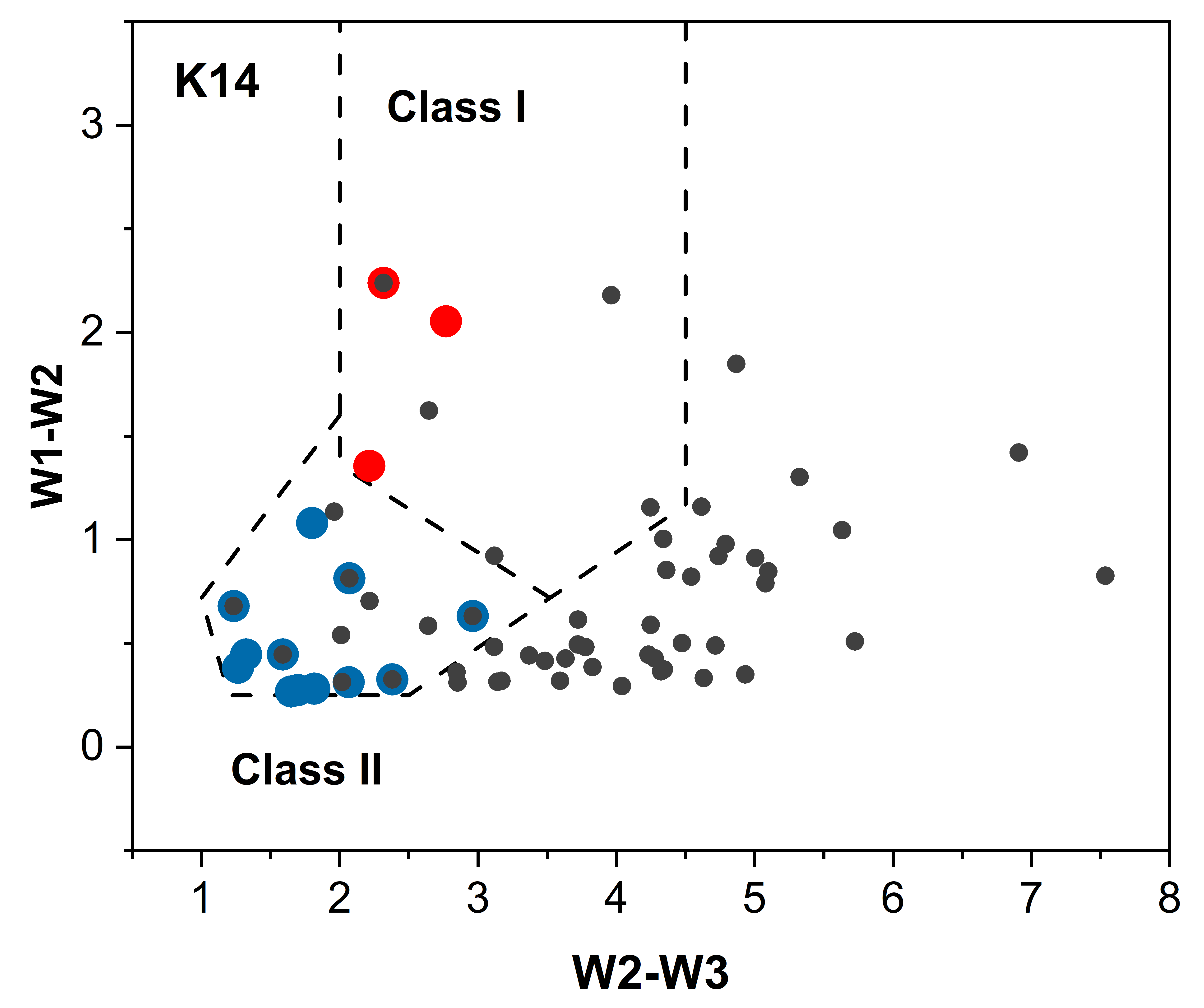}
\caption{Colour–colour diagrams of YSO candidates within the GRSMC\,045.49+00.04 molecular cloud. \textit{Top left panel}: (J–H) vs. (H–K) diagram, with black curves showing the loci of dwarfs and giants from \citet{Bessell1988} and converted to the 2MASS photometric system \citep{Carpenter2001}. The parallel lines represent the interstellar reddening vectors \citep{Rieke1985}. The remaining panels are: \textit{top right} - K–[3.6] vs. [3.6]–[4.5]; \textit{middle left} - [3.6]–[4.5] vs. [5.8]–[8.0]; and \textit{middle right} - [3.6]–[5.8] vs. [8.0]–[24], where the Class I and Class II YSO domains are adopted from \citet{Allen2007, Megeath2004, Muzerolle2004}. The bottom panels present W1–W2 vs. W2–W3 diagrams, constructed according to the criteria of \citet{Koenig2012} (\textit{left}) and \citet{Koenig2014} (\textit{right}). The reddening vectors for Spitzer and WISE bands are obtained from \citet{Flaherty2007} and \citet{Xue2016}, respectively. Class I YSOs are indicated by red circles, Class I/II by orange, and Class II by blue. Circle sizes vary to enhance visualization. In the \textit{bottom right panel}, dark gray circles indicate YSO candidates identified by \citet{Koenig2012} and supported by additional c–c diagrams.} 
	\label{fig:c-c}
\end{figure*}

\section{Scientific justification for used data}
\label{Science}
To identify potential YSOs within star-forming regions, we employed a hybrid approach based on both NIR and MIR photometric data. This combination allows for a comprehensive analysis of the spectral properties of stars, crucial for detecting the characteristic IR excess indicative of YSOs. By integrating data across these spectral ranges, we enhance the reliability of our identifications, accommodating the varied environments and stages of star formation present in these regions.

\subsection{Near-infrared}
\label{NIR}

We utilized JHK magnitudes from the UKIRT Infrared Deep Sky Survey Galactic Plane Survey DR6 \citep[UKIDSS\,GPS; ][]{Lucas2008}, the averaged JHKs photometry from multiple epochs in the VISTA Variables in the Vía Láctea survey \citep[VVV;][]{Minniti2010}, as provided by the VVV Infrared Astrometric Catalogue \citep[VIRAC;][]{Smith2018}, and JHKs magnitudes from the Two Micron All Sky Survey \citep[2MASS;][]{Skrutskie2006}. The UKIDSS\,GPS is designed to survey the equatorial Galactic plane, primarily targeting star-forming regions within Galactic latitudes of -5° to +5°. The VVV survey complements this by covering the southern Galactic plane and the central bulge region. Although the 2MASS survey has a lower photometric sensitivity and resolution compared to UKIDSS\,GPS and VIRAC, it remains one of the few surveys offering a uniform, all-sky catalogue of celestial objects in the NIR. In addition, 2MASS provides high-quality, well-calibrated astrometric and photometric data, establishing a fundamental point of comparison for the brightness and position of IR sources.

\cite{Hernandez2005} and \cite{Meyer1997} found that Class II middle-mass Herbig Ae/Be stars and low-mass classical TTauri (TTau) stars exhibit significant NIR excess emission arising from optically thick dusty inner disks. Conversely, the intrinsic NIR colours of weak TTau objects (Class III) align closely with those of normal MS dwarf stars. Earlier, \cite{Lada1992} showed that Class I YSOs in the (J-H) vs. (H-K) colour-colour (c-c) diagram (see Fig. \ref{fig:c-c}) are found in a region characterized by high extinction values (J-K \textgreater\,3). Moreover, it was shown that there is an inverse relationship between the frequency of NIR excess emission and the age of the cluster \citep{Briceno2007}, highlighting the utility of the JHK c-c diagram as a powerful tool for distinguishing stellar objects with IR excess. However, when identifying stellar objects with an IR excess, it is essential to consider that deviations from the MS may arise not only from the presence of circumstellar disks and envelopes, but also from interstellar extinction along the reddening vectors. The IR excess of sources located to the right of these vectors cannot be attributed solely to interstellar extinction; it also partially results from circumstellar matter. Therefore, objects to the right of the reddening vectors can be considered YSO candidates with a reasonable degree of certainty, except for classical Be stars, which may have smaller IR excesses (J–K \textless\,0.6 and H-K \textless\,0.3) but still reside to the right of the MS reddening band \citep{Hernandez2005}. Among the objects located in the reddening band of MS and giants, those with a (J-K \textgreater\,3) colour index have been classified as Class I YSOs. Although this approach may overlook some objects with relatively minor IR excesses that lie within the reddening band, it ensures a relatively high sample purity. This approach prioritizes the identification of objects with more pronounced IR signatures, which are more likely to be genuine YSOs, thereby minimizing the inclusion of false positives in the sample.

\subsection{Mid-infrared}
\label{MIR}

For the MIR bands, we utilized photometric data obtained from the Spitzer Space Telescope and the Wide-field Infrared Survey Explorer (WISE). These data sources provide comprehensive MIR coverage, which is essential for analyzing the characteristics of YSOs, particularly their warm dust emissions, which are prominent in these wavelengths.

\subsubsection{Spitzer data}

Spitzer has proven to be effective in identifying YSO candidates \citep[e.g.][]{Allen2007, Gutermuth2008}. The selection of YSOs is largely based on the four-channel (3.6\,$\mu$m, 4.5\,$\mu$m, 5.8\,$\mu$m, and 8.0\,$\mu$m) Infrared Array Camera (IRAC) photometry from Galactic Legacy Infrared Mid-Plane Survey Extraordinaire \citep[][and ref. therein]{Churchwell2009}, which include GLIMPSE\,I (31,184,509 sources), GLIMPSE\,II (19,067,533 sources), and GLIMPSE\,3D (20,403,915 sources). Longer-wavelength photometry is available from the Multiband Imaging Photometer for Spitzer Galactic Plane Survey (MIPSGAL), which imaged the 24 and 70\,$\mu$m emission along the inner disk of the Milky Way \citep{Gutermuth2015}.

To identify YSO candidates using Spitzer IRAC and MIPSGAL 24\,$\mu$m colour indices, we employ three c-c diagrams: K-[3.6] vs. [3.6]-[4.5]; [3.6]-[4.5] vs. [5.8]-[8.0]; and [3.6]-[5.8] vs. [8.0]-[24]. Research on models of YSOs' envelopes and circumstellar disks, covering a broad spectrum of parameters, demonstrates that objects at the Class I and II evolutionary stages typically occupy distinct regions in the colour space of these diagrams \citep[e.g.][]{Allen2004, Megeath2004}. Consequently, they can be accurately identified using Spitzer photometric data alone, even without additional information such as spectral data. The domains corresponding to Class I and II YSOs in these diagrams, as delineated by \cite{Megeath2004} and \cite{Allen2007} are shown in Fig.~\ref{fig:c-c}.

The diagram in the upper right panel of Fig.~\ref{fig:c-c} bridges the NIR and MIR ranges. IRAC magnitudes are significantly more sensitive at 3.6 and 4.5\,$\mu$m than at 5.8 and 8.0\,$\mu$m; therefore, many sources are detected only at these two shorter wavelengths. On the other hand, not all stellar objects are simultaneously detected in the J, H, and K bands. The K band, which is the least affected by interstellar extinction, typically identifies the largest number of objects. Consequently, this diagram has proven to be effective in identifying a significant number of YSO candidates \citep{Azatyan2022}. In the diagram, empirical diagonal lines outline the region where most of the classifiable YSOs are located. The [3.6]–[4.5] colour is dominated by the accretion rate; therefore, Class I sources with higher accretion rates appear to be redder. The empirical line $(K-[3.6]) > -2.85714 \times (([3.6]-[4.5]) - 0.401) + 1.7$ \citep{Gutermuth2008} separates the domains of Class I and II objects.

The [5.8]–[8.0] colour is insensitive to reddening and effectively distinguishes between highly reddened background stars with only photospheric emission and young stars with circumstellar matter. Sources centred at ([3.6]–[4.5], [5.8]–[8.0])=(0, 0) correspond to stellar photospheres and Class III diskless PMS stars \citep{Megeath2004}. Based on the range of colours represented by the models and confirmed by data from several young clusters, the domains in the [3.6]-[4.5] vs. [5.8]-[8.0] diagram for Class I and II objects were defined (see the middle left panel in Fig.~\ref{fig:c-c}). Class II objects are within the ranges 0 \textgreater\,[3.6]–[4.5] \textgreater\,0.8 and 0.4 \textgreater\,[5.8]-[8.0] \textgreater\,1.1. Objects for which [5.8]–[8.0] \textgreater\,1.1 are consistent with Class I objects, but with [3.6]–[4.5] \textless\,0.4, which is lower than that predicted by Class I models \citep{Allen2004}, are classified as Class I/II sources. Objects with [3.6]–[4.5] \textgreater\,0.8 and [5.8]–[8.0] \textgreater\,0.2 and/or [3.6]–[4.5] $\geq$ 0.4 and [5.8]–[8.0] \textgreater\,1.1, are classified as protostars, including Class 0 and Class I objects.

MIPS 24\,$\mu$m photometric data provides a longer wavelength baseline for the classification of YSOs, particularly useful for resolving reddening degeneracy between Class I and II. This data is also crucial for identifying optically thin evolved disks, which lack excess emission at shorter wavelengths due to the absence of close-in dust. \cite{Muzerolle2004} delineated the loci for Class I and II in the IRAC/MIPS c-c diagram, presented on the middle right panel of Fig.~\ref{fig:c-c}. Class I YSOs with envelope exhibit strong [8.0]–[24] and [3.6]–[5.8] excesses, occupying the domain where [3.6]–[5.8] \textgreater 1.5 and [8.0]–[24] \textgreater 2.4 in the colour space of the [3.6]-[5.8] vs. [8.0]-[24] diagram. Class II YSOs with optically thick discs show comparatively less [3.6]–[5.8] excess and mostly located in [3.6]–[5.8] \textgreater\,0.3 and [8.0]–[24] \textgreater\,2 domain. Stellar sources clustered around [3.6]–[5.8] = 0 and 0 $\leq$ [8]–[24] $\leq$ 1 likely represent a mixture of pure photospheres and Class III PMS objects.

\quad

The classification methods described above implicitly assume that all objects exhibiting an IR excess are YSOs. IR excess, however, can also be displayed by other galactic and extragalactic sources in the diagrams. Among the stellar objects that can contaminate the sample of YSOs, the most notable are evolved Asymptotic Giant Branch (AGB) stars with [4.5] \textgreater\,7.8\,mag and [8.0]–[24.0] \textless\,2.5 colour indices in the IRAC/MIPS diagram \citep{Robitaille2008}.

There are two primary classes of extragalactic contaminants that can be misidentified as YSOs in MIR bands. The first group includes star-forming galaxies and narrow-line active galactic nuclei (AGNs), which exhibit increasing excesses at 5.8 and 8.0\,$\mu$m due to Polycyclic Aromatic Hydrocarbon (PAH) emission, as identified in the IRAC c-c diagrams \citep{Stern2005}. These objects occupy the colour space defined by the criteria: [3.6]-[5.8] \textless\,1.5 $\times$ ([4.5]-[8.0]-1)/2, [3.6]-[5.8] \textless\,1.5 and [4.5]-[8.0] \textgreater\,1. The second group consists of broad-line AGNs, identifiable using the [4.5] vs. [4.5]-[8.0] colour-magnitude diagram. According to \cite{Gutermuth2008} and \cite{Qiu2008}, broad-line AGNs meet the conditions: [4.5]-[8.0] \textgreater\,0.5, [4.5] \textgreater\,13.5+([4.5]–[8.0]-2.3)/0.4, and [4.5] \textgreater\,13.5. 

To ensure the purity of our sample, we excluded all of the above objects from the final list of YSOs.

\subsubsection{ALLWISE data}
Although WISE is less sensitive than Spitzer and has lower spatial resolution, its photometric data still provide valuable insights into the stellar content of star-forming regions and can confirm results obtained using other instruments. The WISE bands (3.4\,$\mu$m (W1), 4.6\,$\mu$m (W2), 12\,$\mu$m (W3), and 22\,$\mu$m (W4)) closely match those of Spitzer, enabling the use of similar methods to identify YSOs. We used data from the ALLWISE catalogue, which is generally about 0.2 to 0.3 magnitudes more sensitive than the original WISE catalogue for most of its detections \citep{Wright2010}. 

A widely used approach to identify YSOs involves c-c diagrams to isolate regions of IR excess that signify stellar youth. This technique has been thoroughly demonstrated by \citet{Koenig2012} and \citet{Koenig2014} using WISE W1–W2 and W2–W3 color indices. The latter version introduces more stringent photometric quality criteria and a multi-step procedure for removing both stellar and extragalactic contaminants. Although these refinements improve the purity of the selected sample, they may also lead to the exclusion of some genuine YSOs, particularly those that are faint or lie near the classification boundaries. As will be shown in Sec.~\ref{results}, this can result in a significant reduction in the number of detected YSOs. To ensure flexibility in the analysis, we provide the option to apply either selection method at the user's discretion.

According to \citet{Koenig2012} Class I YSOs, the reddest objects exhibit W1-W2 \textgreater\,1.0 and W2–W3 \textgreater\, 2.0 (see Fig.~\ref{fig:c-c}). Class II YSOs are slightly less red, characterised by W1–W2 - $\sigma$(W1–W2) \textgreater\,0.25 and W2–W3- $\sigma$(W2–W3) \textgreater\,1.0, where $\sigma$(...) represents the combined photometric error added in quadrature. YSO candidates are further verified using the W4 band. Class I sources are reclassified as Class II if their W2-W4 \textless\,4.0, and Class II stars are returned to the unclassified pool if their W1–W3 \textless\,-1.7 $\times$ (W3–W4) + 4.3. 

According to \citet{Koenig2014}, sources are classified as Class I YSOs if their colors satisfy the following criteria:
W2-W3 \textgreater\,2.0, and
W1-W2 \textgreater\,-0.42 × (W2-W3) + 2.2, and
W1-W2 \textgreater\,0.46 × (W2-W3) - 0.9, and
W2-W3  \textless\,4.5.  Class II YSOs, including candidate T\,Tauri and Herbig Ae/Be stars, occupy a region defined by the following colour criteria: W1-W2 \textgreater\,0.25 and W1-W2 \textless\,0.9 × (W2-W3) - 0.25, and W1-W2 \textgreater\,-1.5 × (W2-W3) + 2.1, and W1-W2 \textgreater\,0.46 × (W2-W3) - 0.9, and W2-W3  \textless\,4.5. Contamination from extragalactic sources and evolved stars was evaluated based on W4-band photometry.

\begin{figure*}
	\centering 
	\includegraphics[width=1.0\textwidth, angle=0]{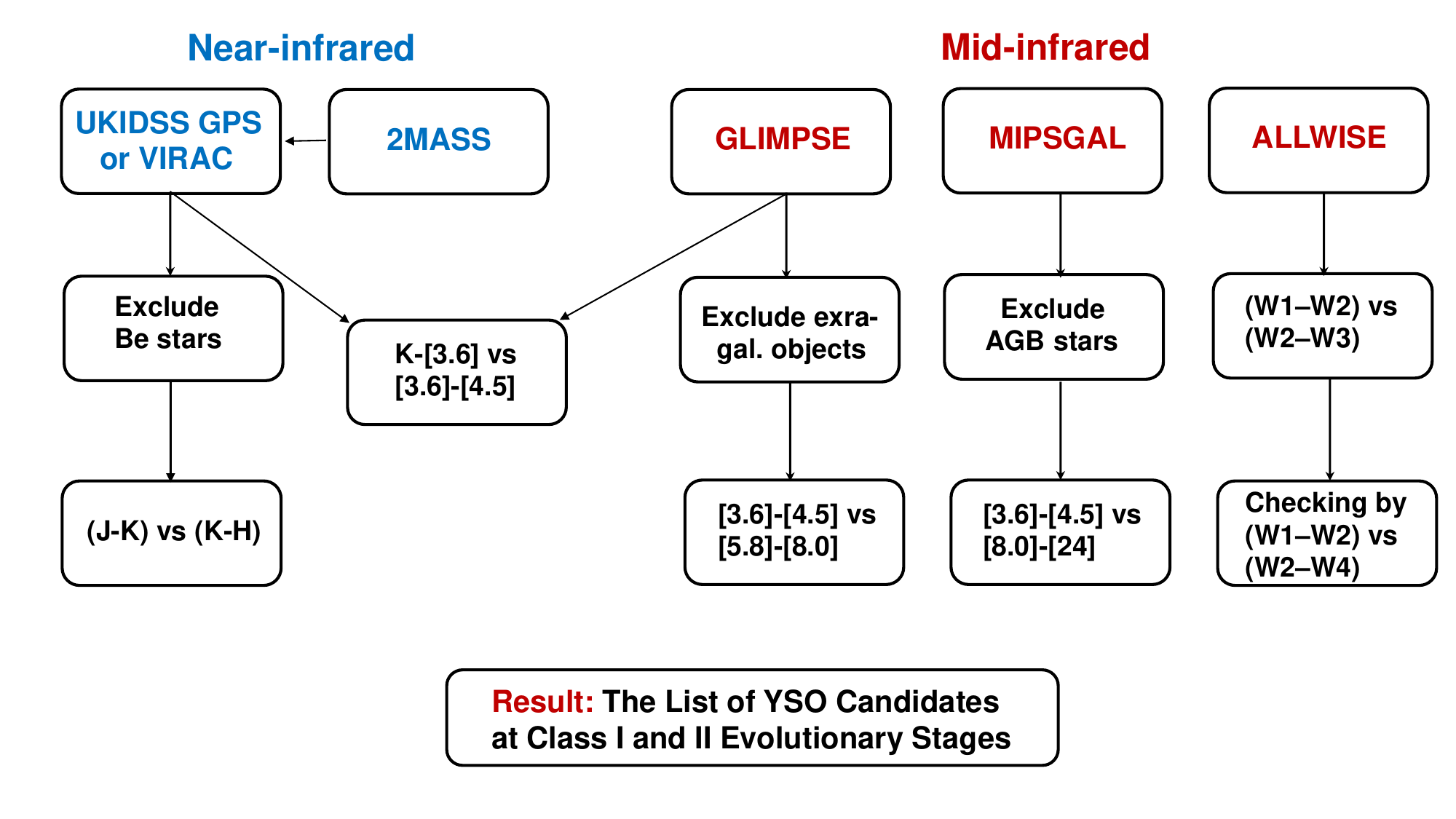}	
	\caption{Schematic representation of the \texttt{IdentYS} code} 
	\label{fig:code}
\end{figure*}

\section{Python code}
\label{code}

The schematic representation of the \texttt{IdentYS} code is illustrated in Fig.~\ref{fig:code}, with an explanation of its operating principles provided below. All photometric data, restricted by specific criteria to ensure sample purity, are utilized from the VizieR database.

We have implemented several restrictions in the UKIDSS\,GPS-DR6 catalogue based on the characteristics described in \citet{Lucas2008}. The catalogue assesses the likelihood of an object being a star, galaxy, or noise by analysing its image profile. We select all objects with less than a 33\% probability of being noise and retain all objects identified as "galaxies" and "probable galaxies". Given that YSOs are often embedded within nebulae, they may be classified as extended nebulous objects; excluding these would inevitably result in the loss of stellar data. On the other hand, the significant interstellar extinction typical of star-forming regions makes the presence of projected extragalactic objects unlikely. We also considered the photometric limits in all three bands and excluded objects with J\textgreater19.77, H\textgreater19.00, and K\textgreater18.05 magnitudes. In addition, objects with zero error measurements in the J, H, and K bands have been excluded from the catalogue. Furthermore, we excluded objects whose K-band quality flag (k\_1ppErrBits) exceeds 64, as this suggests contamination by artifacts such as nebulous structures, diffraction spikes from bright stars, or sources near the detector boundary. As noted in \cite{Lucas2008}, objects with J\textless13.25, H\textless12.75, and K\textless12.0 magnitudes are saturated. Using the correlation established in \cite{Carpenter2001}, we convert the UKIDSS magnitudes to the 2MASS photometric system. Where available, saturated magnitudes were replaced with 2MASS values. 

The Python code also supports the use of data from the VVV\,VIRAC catalogue \citep{Smith2018}. These NIR data were restricted by photometric limits: J\textgreater19.5, H\textgreater18.6, and Ks\textgreater18.00 magnitudes.  Sources with zero photometric errors in the J, H, and Ks bands have been excluded from the catalogue. We also excluded objects with a high probability of being noise (i.e., "mergedClass"\,=\,-1) and those with a quality flag "Jperrbits" \textgreater 256.  Transformations to the 2MASS photometric system were applied using the relations from \citet{Gonzalez2018}. Where available, the saturated VVV magnitudes \citep[J\textless9.5, H\textless10.0, and K\textless10.5;][]{Saito2012} were replaced with the corresponding 2MASS values.

From the GLIMPSE catalogue, we utilize only high-reliability sources with a signal-to-noise ratio greater than 5, designated with a quality flag of "C". From ALLWISE, following \citet{Koenig2012}, we select objects with good photometric detections, specifically those with photometric uncertainties less than 0.2\,mag in all bands. In the later version \citep{Koenig2014} the selection criteria are more stringent. They require that sources have non-null profile-fit magnitude uncertainties (sigmpro), and apply signal-to-noise (\texttt{snr}) and reduced $\chi^2$ thresholds as follows:
\begin{itemize} 
 \item for W1 mag $\chi^2$ \textless (\texttt{snr}-3)/7;
 \item for W3 mag \texttt{snr}\,$\geq$\,5 and either $\chi^2$ \textless (\texttt{snr} - 8)/8 or 0.45 \textless $\chi^2$ \textless 1.15; 
 \item for W4 mag $\chi^2$ \textless (2 × \texttt{snr} - 20)/10 
\end{itemize}.

The purity of the YSO sample was enhanced by excluding field contamination from both stellar and extragalactic sources, as described in Sec.~\ref{Science}. Only after this filtering were stars with IR excess identified based on their colour indices in the diagnostic diagrams.

\begin{table}
\centering 
\begingroup
\small
\setlength{\tabcolsep}{12pt}
\begin{tabular}{l c l } 
 \hline
 Catalogue name & Positional  & Reference \\ 
 & accuracy ($^{\prime\prime}$) & \\
 \hline
2MASS & 0.1 & \cite{Skrutskie2006}\\
UKIDSS\,GPS & 0.3 & \cite{Lucas2008} \\
VIRAC & \textless 0.1 & \cite{Smith2018} \\
GLIMPSE & 0.3 & \cite{Churchwell2009} \\
MIPSGAL & 1.0 & \cite{Carey2009} \\
ALLWISE & 1.0 & \cite{Wright2010}\\
 \hline
\end{tabular}
\caption{Positional accuracy of databases}
\label{Table1}
\endgroup
\end{table}

\begin{figure}[h]
	\centering 
	\includegraphics[width=0.48\textwidth, angle=0]{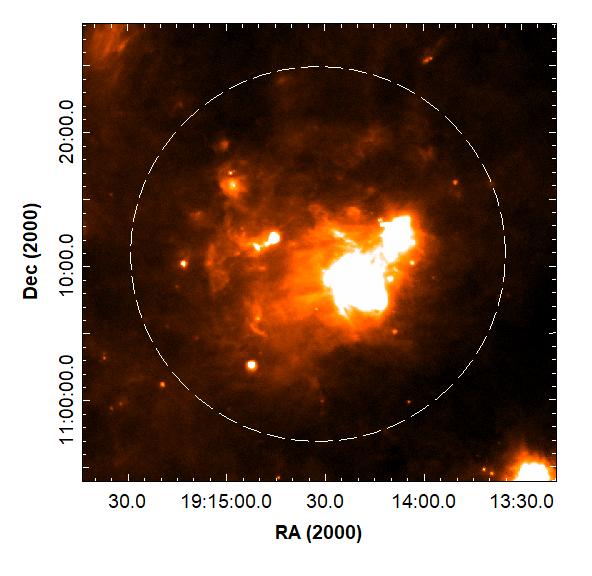}	
	\caption{\textit{Herschel} 70\,$\mu$m image of the GRSMC\,045.49+00.04 molecular cloud. The dashed white circle highlights an area with a 14$^{\prime}$ centered around the coordinates 19:14:32.0, +11:10:55.0, delineating the primary study region.} 
	\label{fig:image}
\end{figure}

Objects from various catalogues were cross-matched within a 3$\sigma$ matching radius, accounting for the combined positional uncertainties of the databases (see Table \ref{Table1}). Our code performs this cross-matching independently, without prioritising any catalogue as the primary reference. The procedure is conducted in multiple stages. First, the coordinates of UKIDSS\,GPS or VIRAC sources are cross-matched with those from GLIMPSE, retaining all objects regardless of whether they are detected in one or both catalogues. The coordinates of MIPSGAL sources not identified in GLIMPSE are then compared with those of UKIDSS\,GPS/VIRAC, and all matches are retained. Finally, ALLWISE sources are cross-matched with UKIDSS\,GPS/VIRAC, and, if no counterpart is found, the matching is performed sequentially with GLIMPSE and then MIPSGAL.

The higher spatial resolution of the NIR catalogues allows for the potential association of multiple NIR sources with a single IRAC in IRAC+UKIDSS matching. However, \citet{Morales2017} reported that in IRAC+UKIDSS matching, the NIR flux is typically dominated by a single counterpart. Therefore, in cases where multiple NIR sources are present, preference is given to the brightest object in the K-band within the cross-matched radius. The same approach was applied for the cross-correlation between NIR and ALLWISE sources, as well as MIPSGAL sources.

As a result, we will compile a list of YSO candidates, including:
\begin{itemize}
    \item Source designation;
    \item Astrometric data (RA, Dec, and projected distance in arcmin from searching center);
    \item Photometric data (magnitudes and associated errors from UKIDSS\,GPS or VIRAC converted to 2MASS photometric system, as well as IRAC, MIPS, and ALLWISE measurements);
    \item Evolutionary stages, determined by the presence of IR excess by specific c-c diagrams: NIR ((J-H) vs. (H-K)), MIR1 ([3.6]-[4.5] vs. [5.8]-[8.0]), MIR2 ([3.6]-[4.5] vs. [8.0]-[24]), NMIR (K-[3.6] vs. [3.6]-[4.5]), and W ((W1-W2) vs. (W2-W3);
    \item Average evolutionary stages for YSO candidates, with the final classification (Class I or II) based on the majority result. In cases of equal weighting, the source will be classified as intermediate Class I/II.
\end{itemize}
Additionally, a secondary file will provide the designation, astrometric, and photometric parameters for sources within the search field that do not meet the YSO criteria, enabling further post-analysis.

Unfortunately, UKIDSS\,GPS and VVV\,VIRAC surveys do not provide full-sky coverage. Consequently, our code is designed to use data exclusively from the 2MASS survey when required. In such case, the final list of YSO candidates contains 2MASS photometric measurements. Additionally, the code supports simultaneous searches for YSO candidates across multiple regions. The complete \textit{source code}, including documentation and example datasets, is available at \url{https://github.com/DanielBaghdasaryan/identys}. 

\section{Results and discussion}
\label{results}

\subsection{Colour-Colour diagrams}

To demonstrate the operation of the proposed \texttt{IdentYS} code, we present its application to the GRSMC\,045.49+00.04 molecular cloud - an active, distant \citep[$\sim$ 8\,kpc; ][]{Wu2019}, and deeply embedded \citep[A$_v$ = 11\,mag; ][]{Azatyan2024} star-forming region (see Fig.~\ref{fig:image}). Fig.~\ref{fig:c-c} shows the results of YSO candidate selection within this molecular cloud, encompassing a 14$^{\prime}$ radius centred around the coordinates 19:14:32.0, +11:10:55.0.

In total, across three databases - UKIDSS\,GPS-DR6, Spitzer GLIMPSE, MIPSGAL, and ALLWISE - within the area, we identified over 140,000 point sources that meet the selection criteria described in Sec. \ref{code}. Of these point sources, $\sim$\,19,000 objects exhibit an IR excess on at least one c-c diagram. However, for the purity of the final sample, relying solely on the results from one diagram when selecting membership within a star-forming region is not advisable. This caution is due to various factors, including errors in the database, unresolved binary objects, and the inclination of the disk component relative to the line of sight, among others. To minimize the risk of incorrect identification, we selected YSO candidates based on the criterion that they must be classified as exhibiting IR excess in at least two different c-c diagrams. YSO candidates identified solely from the W1-W2 vs. W2-W3 diagram were included in the final list, as their youth was also confirmed by the W1-W3 and W3-W4 colour indices. The selection of objects based on at least two diagrams significantly reduced their number to $\sim$\,2,000. For comparison, within the same area, the SPICY catalogue identified 78 YSO candidates with assigned evolutionary classes and 1,082 objects that exhibit strong variability \citep{Kuhn2021}.

The six diagrams in Fig.\,\ref{fig:c-c} illustrate the positions of the selected YSO candidates, with their evolutionary stages determined primarily by the majority of diagrams. In the case of equality, the object was classified as an intermediate Class I/II. In total, 111 Class I, 1,469 Class I/II, and 376 Class II were identified. It is recognized that the dominant radiation from the circumstellar disk varies with distance from the central star, transitioning from the NIR range in the inner regions closest to the star to the FIR range in the more distant outer regions \citep{Hartmann2009}. Consequently, IR excess measurements in the NIR or MIR are generally influenced by the inclination of the disk component relative to the line of sight, which may account for the significant number of Class I/II objects.

Several insights can be drawn regarding the distribution of YSOs across different evolutionary classes in the diagrams. Among the $\sim$\,2,000 selected YSO candidates, only 371 ($\sim$\,18\%) with corresponding photometric data lie outside the expected YSO domains in the classification diagrams. Of these, 73 are located between the reddening vectors in the NIR diagram with J-K$<$3, which does not provide a strong justification for their classification as YSO. Therefore, MIR colour indices have proven effective in identifying numerous YSOs located within the reddening band of the NIR diagram. Conversely, the combined use of NIR and NMIR diagrams has enabled the detection of YSOs that would not have been identified on the basis of MIR photometric data alone. Notably, with only a few exceptions, objects exhibiting the strongest IR excess across all diagrams are predominantly found within the Class I domain. 

The bottom left and right panels present the (W1–W2) vs. (W2–W3) diagrams constructed using the criteria described by \citet{Koenig2012} (K12) and \citet{Koenig2014} (K14), respectively. According to K14 criteria, only 15 YSO candidates were identified: 3 Class I (red circles) and 12 Class II (blue circles). In contrast, the K12 method yielded a significantly larger number of candidates (148), of which 54, marked by a gray circle in the bottom right diagram, are also identified as YSOs in other diagrams, including MIR1 and MIR2. Among these, 20 are located within the Class I and Class II selection boxes. This discrepancy can be attributed to the stricter selection criteria applied in K14, particularly those based on the $\chi^2$ and \texttt{snr}. The remaining YSO candidates are mostly located within the regions occupied by spiral (Sbc) and irregular (Im) galaxies, as shown in Fig. 4 of K14. Although these quantitative ratios may vary between different star-forming regions, it can be concluded that the criteria from K14 yield a significantly cleaner sample of YSOs, albeit at the expense of potentially excluding genuine sources. Therefore, the code provides the ability for the user to choose between these two criteria.

Among the $\sim$\,17,000 objects that exhibit an IR excess in a single c-c diagram, the majority (80\%) were identified as YSO candidates based solely on their NIR photometric data. Further 17\% were classified using the K-[3.6] vs. [3.6]-[4.5] diagram, while only 3\% were identified exclusively from the MIR photometric data. This distribution is likely driven by the relative availability of objects in each database. In the initial data set, which includes $\sim$\, 1,140,000 sources $\sim$\,60\% were detected in the JHK bands, 12\% at 4.5\,$\mu$m, 3\% at 8.0\,$\mu$m, and only 0.6\% in ALLWISE. Approximately 40\% of these 17,000 objects were identified in both the NIR and MIR ranges; however, their IR excess was detected in a single diagram. Within the remaining 60\%, additional YSO candidates may be present if supplementary photometric data become available. Nevertheless, if ensuring sample purity is a priority, selecting YSOs on the basis of a single diagram may introduce contamination into the final sample. Ultimately, the criteria for identifying YSO candidates depend on both the specific research objectives and the chosen methodology. For example, in ultracompact H\,II regions, where saturation is frequently encountered in the MIR range \citep{Fuente2020}, only NIR photometric data, and in very rare cases FIR data, can be used to identify YSOs \citep{Azatyan2022,Azatyan2024}. 

\begin{figure}[h]
	\centering 
	\includegraphics[width=0.44\textwidth, angle=0]{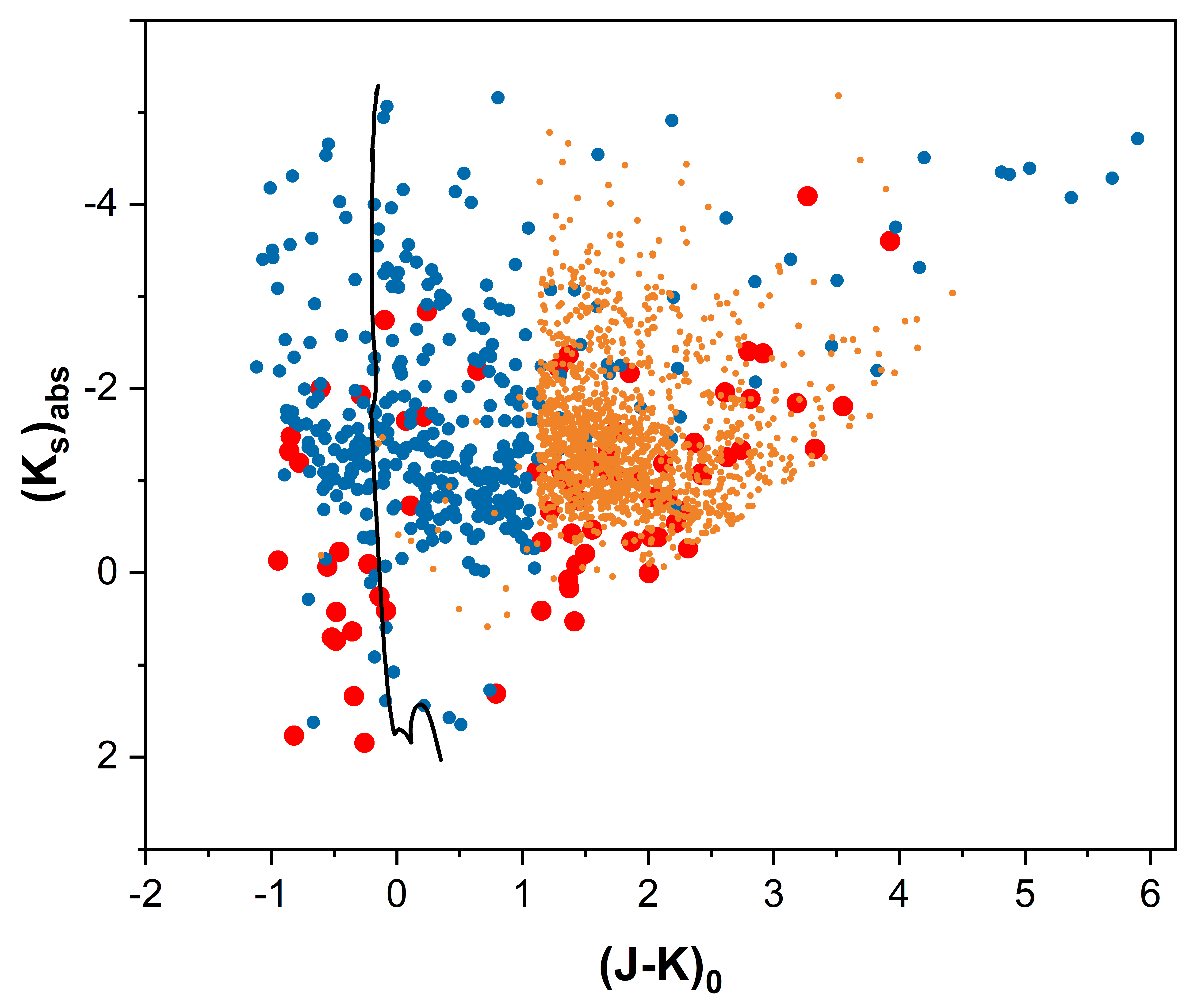}	
 \includegraphics[width=0.44\textwidth, angle=0]{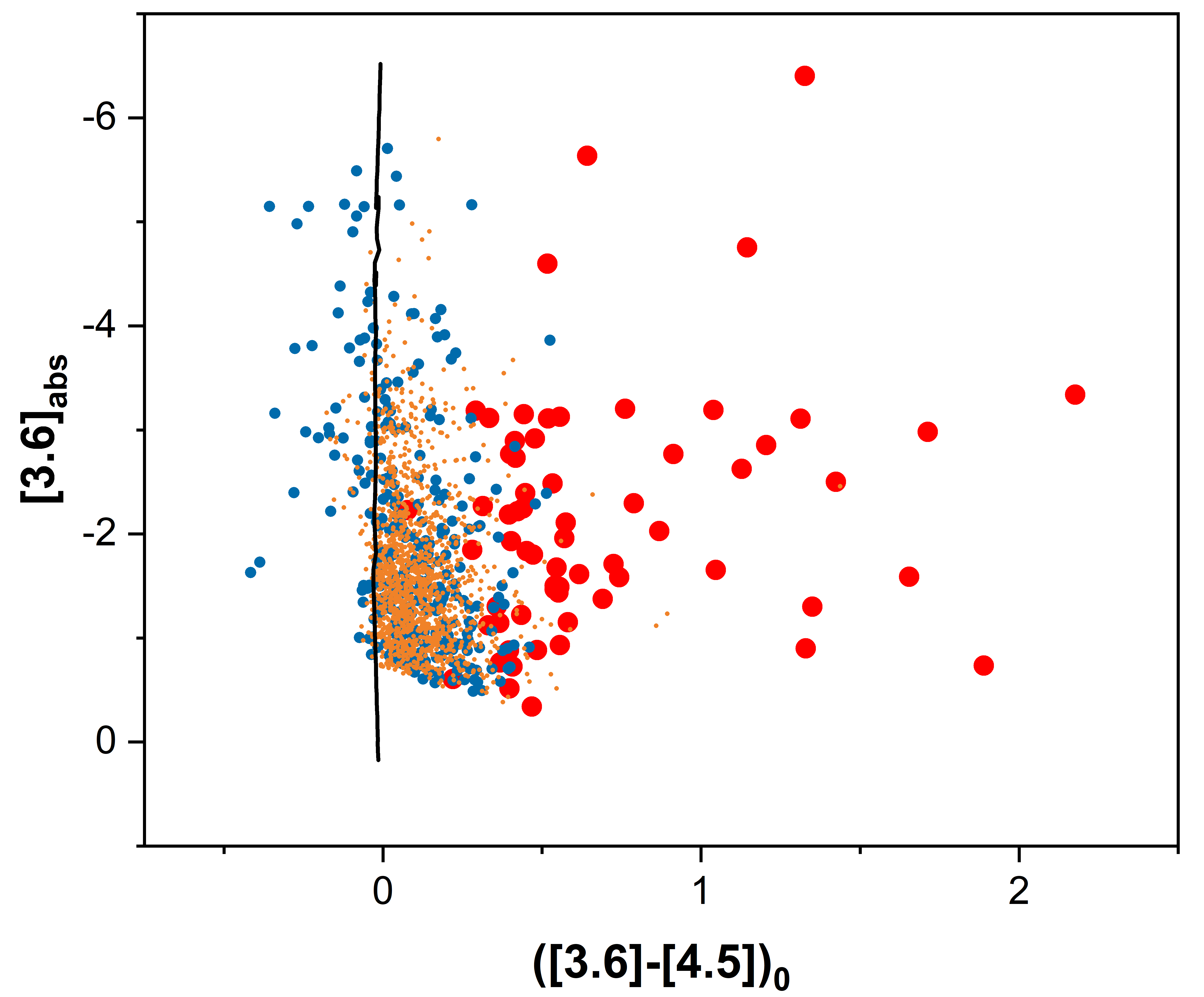}	
  \includegraphics[width=0.44\textwidth, angle=0]{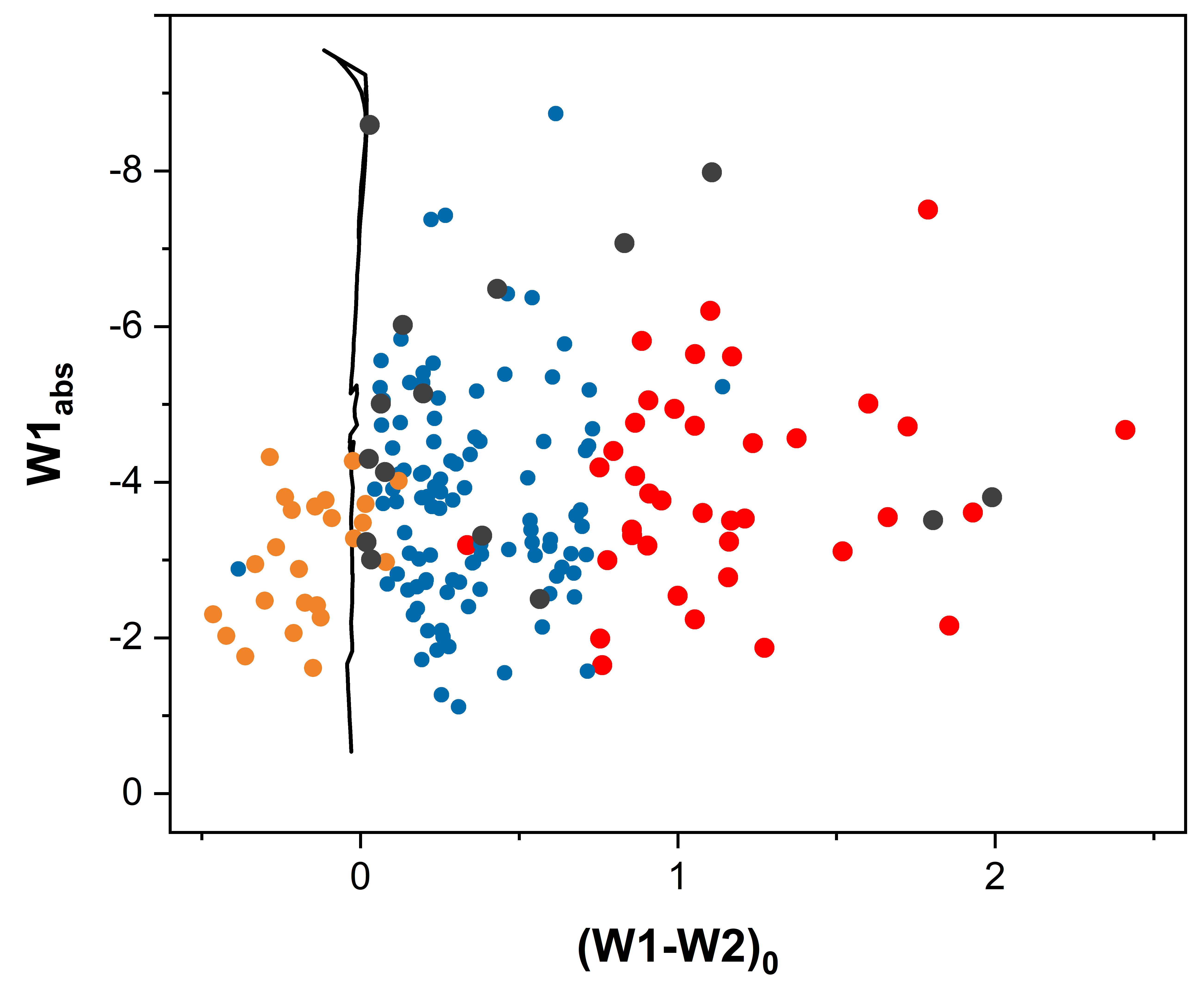}	
 
\caption{Colour–magnitude diagrams of YSO candidates within the GRSMC 045.49+00.04 molecular cloud: 2MASS (\textit{top panel}), Spitzer (\textit{middle panel}), and WISE (\textit{bottom panel}). The black curve represents the 10$^7$ years isochrone adopted from the PARSEC database. The symbols denoting YSOs of different evolutionary classes are the same as in Fig. \ref{fig:c-c}.}
	\label{fig:CMD}
\end{figure}

\begin{figure*}[h]
	\centering 
	\includegraphics[width=0.49\textwidth, angle=0]{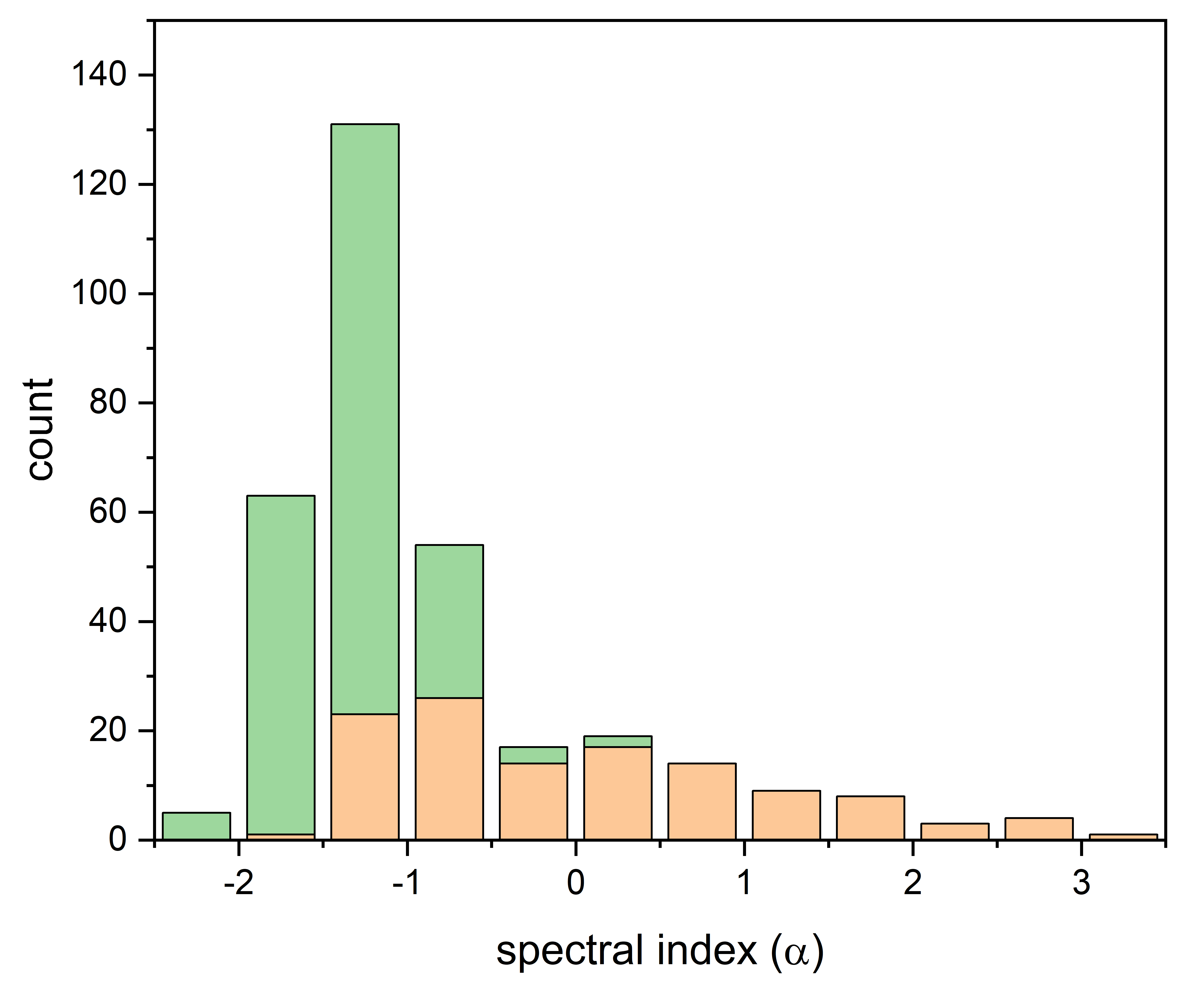}	
 \includegraphics[width=0.5\textwidth, angle=0]{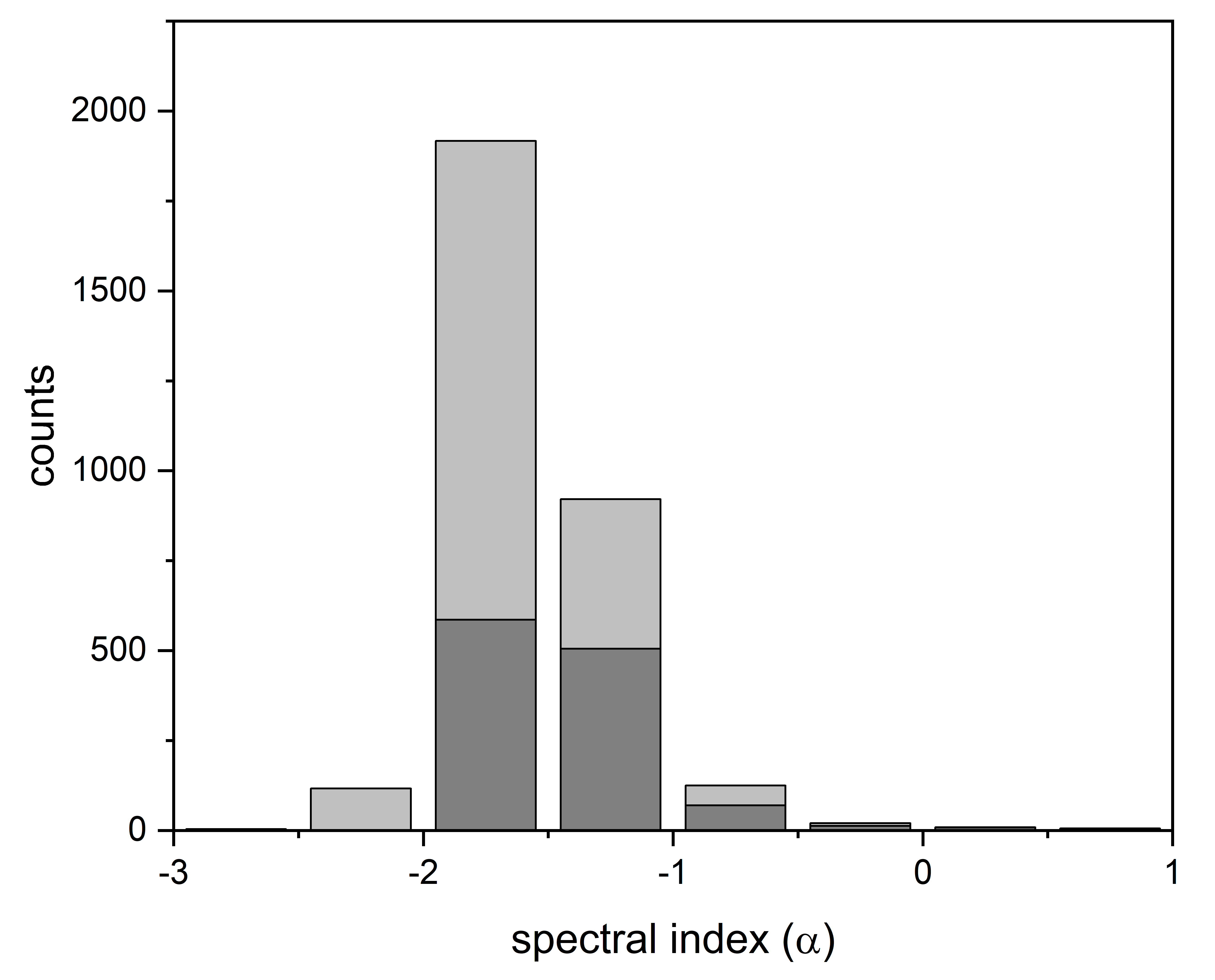}	
 
\caption{Distribution of spectral index $\alpha$. \textit{Left panel}: YSO candidates, for which an IR excess was detected (beije) and not detected (green) in the MIR1 ([3.6] - [4.5] vs. [5.8] - [8.0]) diagram. \textit{Right panel}: stellar objects with IR excess detected in only one diagram (dark grey) and those with no IR excess detected (light grey).}
	\label{fig:SED}
\end{figure*}

\subsection{Colour-magnitude diagram}

In addition to c–c diagrams, various other techniques can be employed to identify YSO candidates. One such method is the colour–magnitude diagram (CMD), which provides valuable insight into stellar evolutionary stages by comparing the positions of sources with theoretical isochrones. We further examined our sample of YSO candidates - selected based on at least two c-c diagrams - using a CMD. Specifically, we evaluated their positions relative to the isochrone corresponding to an evolutionary age of 10$^7$ years, which can reasonably be considered representative of the zero-age main sequence (ZAMS) for low-mass stellar objects \citep{Stahler2004}. The isochrone was obtained from the PARSEC database \citep[Padova and Trieste Stellar Evolution Code;][]{Bressan2012} through the CMD 3.8 web interface (\url{ttp://stev.oapd.inaf.it/cgi-bin/cmd}). 

To include all objects in the YSO candidates sample, three CMDs were constructed: (K$_s$)$_{abs}$\,vs.\,(J-K)$_0$, [3.6]$_{abs}$\,vs.\,([3.6]-[4.5])$_0$, and W1$_{abs}$\,vs.\,(W1-W2)$_0$. These CMDs are shown in Fig. \ref{fig:CMD}. The apparent magnitudes of the YSO candidates were converted to absolute magnitudes using a distance of D = 7.8\,kpc and an interstellar extinction of A$_v$ = 11 \,mag, adopted from \citet{Azatyan2024}. Interstellar extinction corrections were applied using the reddening coefficients provided by the CMD 3.8 web interface (PARSEC).

The diagrams clearly demonstrate that the overwhelming majority of stellar objects exhibit a significant displacement to the right of the 10$^7$ years isochrone, supporting their classification as YSOs. Furthermore, in both MIR diagrams (middle and bottom panels), all Class I objects exhibit the greatest rightward offsets from the isochrone. It should also be noted that all objects identified using the K14 method (see Sec. \ref{code}) lie to the right of the isochrone. In general, only about 10\% of the YSO candidates are located to the left of the isochrone: 8\% in the 2MASS CMD, 10\% in the Spitzer CMD and 12\% in the WISE CMD. This fraction can be explained both by uncertainties in object identification and by errors in distance determination, as well as by a significant gradient of interstellar extinction in the region.

\subsection{Spectral energy distribution}

The IR spectral index, defined as $ \alpha = \frac{dlog(\lambda f_{\lambda})}{dlog(\lambda)}$, is also frequently used to identify YSOs and classify their evolutionary stages \citep[e.g., ][]{Lada1987}. Calculating this index is most effective when using a broad spectral range that encompasses radiation from the central stellar object, the circumstellar disk, and the envelope, as the relative contributions of these components largely determine the evolutionary stage. Interstellar extinction, however, can significantly alter the slope of the spectral energy distribution (SED), thereby affecting the derived spectral index.

When selecting a spectral range for comparative analysis of different samples, preference is generally given to the range that includes the largest number of identified objects. In our initial sample, fewer than 10\% were identified in the MIR range, with the majority detected only in the NIR, which is insufficient for accurately classifying their SEDs. Nevertheless, we compared the spectral indices $\alpha$ in the Spitzer/IRAC MIR range (from 3.6 to 8.0\,$\mu$m) across three subsamples: YSO candidates with an IR excess detected in at least two diagrams (328 YSO candidates), those with an IR excess detected in only one diagram  (1,228 objects) and those without detected IR excess (1,933 objects). The resulting distributions of the spectral index $\alpha$ for these sub-samples are shown in Fig. \ref{fig:SED}. 

Typically, objects are classified according to their spectral index $\alpha$ values: $\alpha > 0.3 $ -- Class\,I YSOs; $-0.3 \leq \alpha < 0.3 $ -- Class\,I/ II YSOs with a `flat' spectrum; $-1.6 \leq \alpha < - 0.3 $ -- Class II YSOs; and $\alpha < - 1.6 $ -- Class III PMS stars \citep[e.g.][]{Allen2007}. The distribution of $\alpha$ for the YSO candidates closely corresponds to their evolutionary stages, as indicated by their positions on the MIR1 diagram. For stars located within the YSOs domains, $\alpha$ values are predominantly greater than -1.5. YSO candidates selected from other diagrams, but also have photometric data in the 3.6-8.0 range, exhibit shallower slopes, with peak values typically ranging from -1.5 to -1.0. The peak of the $\alpha$ distribution for the other two samples is shifted further to the left, as illustrated in Figure \ref{fig:SED} (right panel), particularly for objects without IR excess detected in any diagram. For the vast majority of these objects, the distribution is consistent with stellar photospheric emission. Unfortunately, only a small number of objects are detected in the 24\,$\mu$m and 22\,$\mu$m bands, which would otherwise help identify PMS stars at the Class\,III evolutionary stage. Although alternative indicators, such as H$_\alpha$ emission, can be used to identify these objects, this method is not suitable for distant and deeply embedded star-forming regions.

\section{Summary}
\label{Summary}
The paper presents the scientific justification and operating principles of the \texttt{IdentYS} tool, designed to identify potential members of star-forming regions and young stellar clusters. The tool facilitates the identification of young stars in distant and embedded star-forming regions, with a primary focus on Class I and II YSOs. It is based on two assumptions: (i) the vast majority of members within active star-forming regions are YSOs; and (ii) the most YSOs located in the direction of the region are likely associated with it. On the other hand, one of the crucial observational indicators of YSOs is the IR excess, caused by the presence of integral structural components of young stars at an early stage of evolution, namely circumstellar envelopes and disks. Accordingly, our approach focuses on detecting stellar objects exhibiting IR excess in the direction of star-forming regions. Since significantly more evolved Class III objects with optically thin circumstellar disks exhibit very little or no IR excess, their identification using the code is not effective.

To identify potential YSOs within star-forming regions, we employed a hybrid approach based on both NIR and MIR photometric data from UKIDSS\,GPS or VVV\,VIRAC, 2MASS, Spitzer GLIMPSE, and ALLWISE. All photometric data, restricted by specific criteria to ensure sample purity, are utilised from the VizieR database. The purity of the YSOs sample is further enhanced by excluding field contamination from stellar and extragalactic objects, including AGB stars, star-forming galaxies, and AGNs. For identification of stellar objects with an IR excess, five NIR and MIR c-c diagrams were used: NIR ((J-H) vs. (H-K)), MIR1 ([3.6]-[4.5] vs. [5.8]-[8.0]), MIR2 ([3.6]-[4.5] vs. [8.0]-[24]), NMIR (K-[3.6] vs. [3.6]-[4.5]), and W ((W1-W2) vs. (W2-W3)). As a result, we compiled a list of YSO candidates containing source designations, astrometric and photometric parameters from all the aforementioned databases, and information on their evolutionary stage, determined by the presence of IR excess as indicated in these diagrams.

To minimize the risk of misidentification caused by factors, such as errors in the database, unresolved binaries, or the inclination of the disk component relative to the line of sight, among others, we propose selecting YSO candidates based on the criterion that they must exhibit IR excess in at least two different c-c diagrams.

The application of this program can significantly streamline the statistical analysis of young stellar populations across diverse star-forming regions, which typically involves handling very large volumes of initial data. Designed for ease of use, it allows researchers to efficiently perform complex analyses without requiring advanced programming skills. By automating the processing and analysis phases, the program not only saves time but also enhances the precision and reliability of the results.

\section*{\small Acknowledgements}
\scriptsize{This work was made possible by a research grant number №\,21AG-1C044 from Science Committee of Ministry of Education, Science, Culture and Sports RA. This research has used the VizieR catalogue access tool, CDS, Strasbourg, France. For this research, we also used the data obtained at UKIRT, which is supported by NASA and operated under an agreement among the University of Hawaii, the University of Arizona, and Lockheed Martin Advanced Technology Center; operations are enabled through the cooperation of the East Asian Observatory. We thank our colleagues in the GLIMPSE and MIPSGAL {\itshape Spitzer} Legacy Surveys. This publication makes use of data products from the Wide-field Infrared Survey Explorer, which is a joint project of the University of California, Los Angeles, and the Jet Propulsion Laboratory/California Institute of Technology, and NEOWISE, which is a project of the Jet Propulsion Laboratory/California Institute of Technology. WISE and NEOWISE are funded by the National Aeronautics and Space Administration.}

\scriptsize
\bibliographystyle{ComBAO}
\nocite{*}
\bibliography{references}

\end{document}